\documentclass[fleqn,10pt]{wlscirep}
\usepackage[utf8]{inputenc}
\usepackage[T1]{fontenc}
\usepackage{color}    
\usepackage{graphicx}
\usepackage{dcolumn}
\usepackage{bm}
\usepackage{subfigure}
\usepackage{amssymb}
\usepackage{multirow}
\usepackage[square,sort,comma,numbers]{natbib}
\usepackage{amsmath}
\graphicspath{{plots/}}
\usepackage{hyperref}
\hypersetup{
    colorlinks = true,
    citecolor = blue
}
\usepackage[utf8]{inputenc}
\usepackage{wrapfig}
\usepackage[T1]{fontenc}
\usepackage[normalem]{ulem}
	
\renewcommand{\vec}[1]{\mathbf{#1}}

\title{On the Relation Between Diffusion and Shear Viscosity in Two-Dimensional Magnetized Yukawa Liquids}

\author[1,2,*]{N.~Kh.~Bastykova}
\author[1,2]{T.~S.~Ramazanov}
\author[1,2]{S.~K.~Kodanova}
\affil[1]{Institute for Experimental and Theoretical Physics, Al-Farabi Kazakh National University, 71 Al-Farabi ave.,
 050040 Almaty, Kazakhstan}
\affil[2]{Institute of Applied Sciences and IT, 40-48 Shashkin str., 050038 Almaty, Kazakhstan}

\affil[*]{kodanova@physics.kz}

\begin{abstract}

We analyzed the relationship between shear viscosity and diffusion in a two-dimensional Yukawa liquid subjected to an external magnetic field, using equilibrium molecular dynamics simulations over a broad range of coupling ($\Gamma$) and magnetization parameters, with the latter covering regimes of weak and strong magnetization. The viscous and diffusive transport properties are analyzed via the Green–Kubo formalism and mean-squared displacement, respectively. We find that the product of the shear viscosity and diffusion coefficient exhibits a nontrivial, non-monotonic dependence on the coupling parameter, deviating significantly from the classical Stokes–Einstein (SE) relation across most of the explored parameter space. In the weak-coupling regime ($\Gamma \lesssim 10$), the dependence of the product of the shear viscosity and diffusion coefficient on the coupling parameter follows a power law $\sim 1/\Gamma^c$ with the parameter $c$ being $c>1$ and dependent on the magnetization degree, indicating a strong breakdown of SE behavior. In contrast, at strong coupling ($60 \lesssim \Gamma \lesssim 120$), the standard SE scaling $\tilde{\eta}\tilde{D}_{\alpha} \sim 1/\Gamma$ is approximately restored, independent of the considered magnetic-field strength. The findings are directly relevant to experiments on quasi-magnetized rotating dusty plasmas and provide a benchmark for theoretical descriptions of transport in magnetized strongly coupled matter.
\end{abstract}

\begin{document}

\flushbottom
\maketitle

\thispagestyle{empty}

\section*{Introduction}

Systems in which particles interact via a screened Coulomb potential are often modeled using the Yukawa pair interaction potential, serving as a versatile model for a wide class of physical systems, including dusty plasmas, colloidal suspensions, and other charged systems \cite{Bonitz_2010, PhysRevE.93.053204, POP_Moldabekov_2015, cpp_moldabekov_62, Moldabekov_pop_2018}. Across a wide range of interaction strengths and screening parameters, the Yukawa model has been widely employed to investigate transport phenomena, phase transitions, and self-induced waves in charged-particle systems. For dusty plasmas and colloidal suspensions, two-dimensional (2D) Yukawa systems are of particular interest, as reduced dimensionality enhances collective effects, long-term correlations, and deviations from classical transport theories \cite{Saigo2002,Donko2002,Hartmann2005,Donko2008,Djienbekov, Ludwig2018, 10.1063/1.4922908, cpp_zhandos_2017}.

Transport coefficients, such as shear viscosity and the diffusion coefficient, are fundamental characteristics that govern momentum and mass transport in a medium. Molecular-dynamics studies have shown that the shear viscosity of Yukawa liquids exhibits a non-monotonic dependence on the coupling parameter $\Gamma$, reflecting the competing influence of interparticle correlations and random thermal motion \cite{Saigo2002,Donko2002,Hartmann2005, BASTYKOVA2025108136}. In 2D systems, the peculiar dependence of the viscosity on the coupling and screening parameters differs from that in 3D systems due to enhanced collective excitations \cite{Donko2008,Liu2008}. 

The diffusion characteristics in the Yukawa systems have also been studied extensively. It has been established that the diffusion coefficient decreases monotonically with increasing coupling strength, reflecting the gradual suppression of particle mobility by strong correlations \cite{Hou2009}. In two dimensions, mean-squared displacement often exhibits anomalous behavior on intermediate time scales measured in experiments, before reaching the hydrodynamic regime, necessitating the use of generalized measures of the diffusion coefficient \cite{Liu2008,Metzler2000, PhysRevE.102.033205}. These observations indicate that in 2D systems, the correlation between viscous and diffusive transports may manifest differently from that in 3D systems, motivating a detailed analysis of their mutual relationship.

A classical connection between viscosity and diffusion is provided by the Stokes--Einstein (SE) relation, which predicts that the product of the diffusion coefficient and shear viscosity remains constant. This relation is known to be valid for simple liquids, and deviations from the SE relation have been reported for strongly coupled Yukawa systems, particularly near structural transitions and in the regime of intermediate to strong coupling \cite{Rosenfeld1977,Pond2011,Costigliola2019}. In particular, using equilibrium molecular dynamics simulations (MD) of a two-dimensional Yukawa system, it was demonstrated that the SE relation breaks down near the disordering transition due to the emergence of collective particle motion and dynamical heterogeneity \cite{PhysRevLett.96.015005}, while it is restored at slightly higher temperatures where normal diffusion is recovered \cite{Pond2011}. These findings highlight the strong sensitivity of the viscosity--diffusion relationship to the microscopic dynamical state of the system.

An additional level of complexity arises in the presence of an external magnetic field. Magnetization strongly modifies particle trajectories via cyclotron motion, which can substantially suppress momentum and mass transport. Previous studies have shown that strong magnetization significantly reduces particle diffusion and strongly affects autocorrelation functions of velocity and stress in strongly coupled Yukawa systems \cite{Dzhumagulova2014, Ott2011, Nosenko2006, Djienbekov2, CTPP2025}.

Despite substantial progress in understanding transport processes in strongly coupled Yukawa systems, a detailed study of the relationship between shear viscosity and diffusion in two-dimensional magnetized Yukawa liquids under strong magnetic fields has not been conducted in prior work. In particular, it remained unclear how an external magnetic field modifies the relation between the viscosity and diffusion coefficients, and whether and how magnetization can alter the SE relation.
In this work, we address this open problem by performing a systematic MD investigation of shear viscosity and diffusion in a two-dimensional magnetized Yukawa system across a wide range of coupling parameters and magnetic field strengths. We show that magnetization gives rise to a qualitatively distinct transport regime in which the viscosity–diffusion product exhibits a magnetic-field-dependent power-law scaling. Our results show that magnetic confinement fundamentally reshapes the relation between viscous and diffusive transport in two dimensions. These findings are directly relevant to magnetized dusty plasmas \cite{Hartmann2}.

\section{Theory and Simulation Details}
\subsection{Shear viscosity}
The viscosity was calculated using the standard Green–Kubo approach applied to equilibrium MD simulation data. In this method, the shear viscosity is obtained from the Green–Kubo expression for the stress autocorrelation function:
\begin{equation}\label{eq:Green-Kubo}
\eta = \frac{1}{Sk_BT}\int_{0}^{\infty} C(t) \,dt, 
\end{equation}
where $S$ is the area of the simulation box and $C(t)=\left<P^{xy}(t)P^{xy}(0)\right>$ is the stress autocorrelation function (SACF) of the particles, $P^{xy}$ is the off-diagonal element of the pressure tensor,
\begin{equation}\label{eq:Stress tensor element}
P^{xy}=\sum_{i=1}^{N}\left[mv_{ix}v_{iy}-\frac{1}{2}\sum_{i\neq j}^{N}\frac{x_{ij}y_{ij}}{r_{ij}}\frac{\partial V(r_{ij})}{\partial r_{ij}}\right],
\end{equation}
where $N$ is the number of particles and $r_{ij} = |\mathbf{r_{i}-r_{j}}|$.

\subsection{Diffusion characteristics}
We consider the mean squared displacement (MSD) of the particles as a measure of the particles' diffusivity in the system. The MSD is computed in a standard way as 
 \begin{equation}\label{eq:MSD}
{\rm MSD}(t)=\left<|\vec{r}(t)-\vec{r}(0)|^2\right>.
 \end{equation}

In general, MSD has a power low dependence on time \cite{PhysRevA.18.2345}: 
 \begin{equation}\label{eq:MSD2}
 {\rm MSD}(t) \propto  t^{\alpha}.
 \end{equation}
 where $\alpha$ is the diffusion exponent.

 On short time scales, the MSD exhibits a quadratic time dependence, ${\rm MSD}(t) \propto t^{2}$, corresponding to the ballistic regime. At long time scales, formally defined as $t \to \infty$, the MSD scales linearly with time, ${\rm MSD}(t) \propto t$, which is characteristic of normal diffusion 
 In experiments on two-dimensional dusty plasmas, however, observing a strictly linear time dependence of the MSD is challenging due to the finite size of the dust particle cluster and the limited measurement duration.

Between the ballistic and normal diffusion regimes, intermediate time scales may exhibit anomalous diffusion, characterized by a power-law dependence ${\rm MSD}(t) \propto t^{\alpha}$, where $1 < \alpha < 2$ corresponds to superdiffusion and $\alpha < 1$ to subdiffusion \cite{PhysRevE.78.026409}.

To characterize  MSD in the anomalous diffusion regime, a generalized diffusion coefficient $D_{\alpha}$ is introduced through $ {\rm MSD}(t) = 4D_{\alpha}t^{\alpha}+b$.
Consequently, depending on the time scale under consideration, the diffusion coefficient $D_{\alpha}$ for a two-dimensional system can be defined following Refs. \cite{Hartmann2, PhysRevE.90.013105}:
\begin{equation}\label{eq:Diff}
 D_{\alpha}= \frac{1}{4} \frac{\partial {\rm MSD}(t)}{\partial t^\alpha}.
\end{equation}

\subsection{Simulation Details}

The numerical study was carried out using equilibrium MD of a two-dimensional system of charged particles. The particles were confined to a planar simulation cell with periodic boundary conditions applied along both spatial directions. The size of the simulation cell was determined by the relation $L/a = \sqrt{\pi N}$, where $a$ denotes the characteristic interparticle spacing and $N$ is the total number of particles. All simulations were performed with $N = 10{,}000$ particles, which provides adequate statistical accuracy.

The temporal evolution of the particle ensemble in the presence of a spatially uniform magnetic field oriented perpendicular to the plane of motion was obtained by numerically integrating the equations of motion using a modified velocity-Verlet algorithm that explicitly incorporates the Lorentz force  \cite{Spreiter1999ClassicalMD}.

The influence of the magnetic field was characterized by the magnetization parameter $\Omega$, defined as the ratio of the cyclotron frequency to the plasma frequency of the particles. The simulations covered the range $0 \leq \Omega \leq 1$. To ensure robust statistical convergence of the calculated transport properties, all reported quantities were averaged over 50 independent MD calculations generated from different initial configurations.

\section{ Results and Discussions} \label{s:2}

\subsection{Shear viscosity}

\begin{figure}[t!]
\centering
\includegraphics[width=1.0\textwidth]{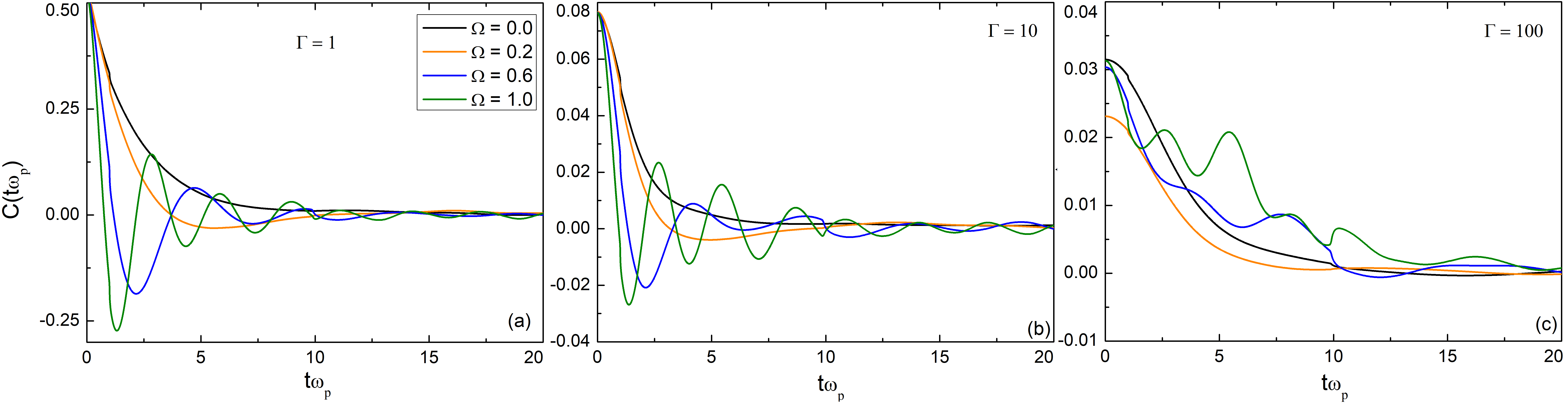}
\caption{  Stress autocorrelation function dependence on time at different values of $\Omega$ for a) $\Gamma = 1$, b) $\Gamma = 10$, and c) $\Gamma = 100$.}
\label{fig:1}
\end{figure}

We first analyze the behavior of the SACF under the influence of an external magnetic field for representative values of the coupling parameter $\Gamma$. The figure \ref{fig:1} shows the SACF as a function of the reduced time $t \omega_p$ for different values of the magnetization parameter $\Omega$ and the coupling parameter $\Gamma$. Three regimes are presented: weak coupling ($\Gamma=1$), intermediate coupling ($\Gamma=10$), and strong coupling ($\Gamma=100$). For each value of $\Gamma$, results are shown for $\Omega = 0.0$, $\Omega =0.2 $, $\Omega = 0.6$, and $\Omega = 1.0$.

\begin{figure}[t!] 
\centering
\includegraphics[width=1.0\textwidth]{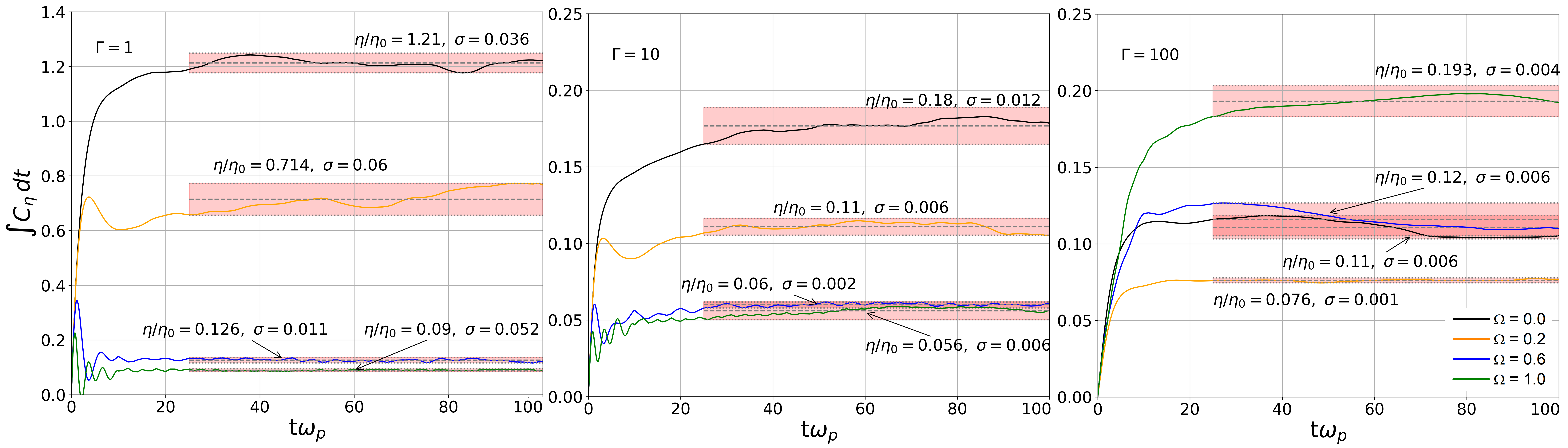}
\caption{Dependence of the integral of the stress autocorrelation function on the upper bound of the integration range at different values of $\Omega$ for a) $\Gamma = 1$, b) $\Gamma = 10$, and c) $\Gamma = 100$.}
\label{fig:2}
\end{figure}

From figure \ref{fig:1}, we see that for $\Gamma=1$, the SACF decays monotonically at $\Omega = 0.0$, and starts to show oscillatory behavior with increasing the magnetization parameter to $\Omega =0.2 $, $\Omega = 0.6$, and $\Omega = 1.0$, indicating a strong influence of the magnetic field. At $\Gamma=10$, the decay of the oscillations with time becomes weaker. As $\Omega$ increases, the relaxation time increases and the oscillatory structure of the SACF becomes more pronounced, indicating a longer persistence of stress correlations in the presence of a magnetic field. For the strongly coupled system $\Gamma=100$, the SACF decays most slowly and remains positive for longer times. Magnetization-induced oscillations are clearly visible in this case as well, although they are no longer around zero. We observe that the magnetic field impact is reflected in enhanced long-time correlations. In general, increasing the coupling parameter $\Gamma$ leads to slower relaxation of the SACF and stronger long-term correlations, while increasing the magnetization parameter $\Omega$ enhances the oscillatory character of the function and modifies its amplitude and decay behavior.

To demonstrate the convergence of the integral in the Green–Kubo relation,  in figure  \ref{fig:2}, we show the dependence of the SACF integral on the upper integration bound at different values of  $\Omega $ for  (a) $\Gamma = 1$, (b) $\Gamma = 10$, and (c) $\Gamma = 100$. Shaded areas indicate the used steady state average intervals, the corresponding normalized viscosity $\eta / \eta_0$ and standard deviations  $\langle \sigma \rangle$ are provided in figure  \ref{fig:2} as well.  For all coupling parameters, the SACF integral rapidly rises at short times ($t \omega_p \lesssim 10$) and reaches a plateau corresponding to the steady-state shear viscosity value. In the weakly coupled regime ($\Gamma = 1$), viscosity strongly decreases with increasing $\Omega$, showing pronounced initial oscillations due to kinetic effects. At intermediate coupling ($\Gamma = 10$), the plateau is lower, and magnetization has a weaker effect on the viscosity value. In the strongly coupled regime ($\Gamma = 100$), the SACF integral reaches a steady value smoothly with small fluctuations, but substantial dependence of the shear viscosity on $\Omega$ remains. Overall, increasing $\Gamma$ accelerates the onset of steady-state with respect to the upper integration bound in the Green–Kubo relation, while increasing $\Omega$ systematically lowers the shear viscosity. 

\begin{figure*}[t!!!]
    \centering
    \begin{minipage}{.5\textwidth}
        \includegraphics[width=\linewidth]{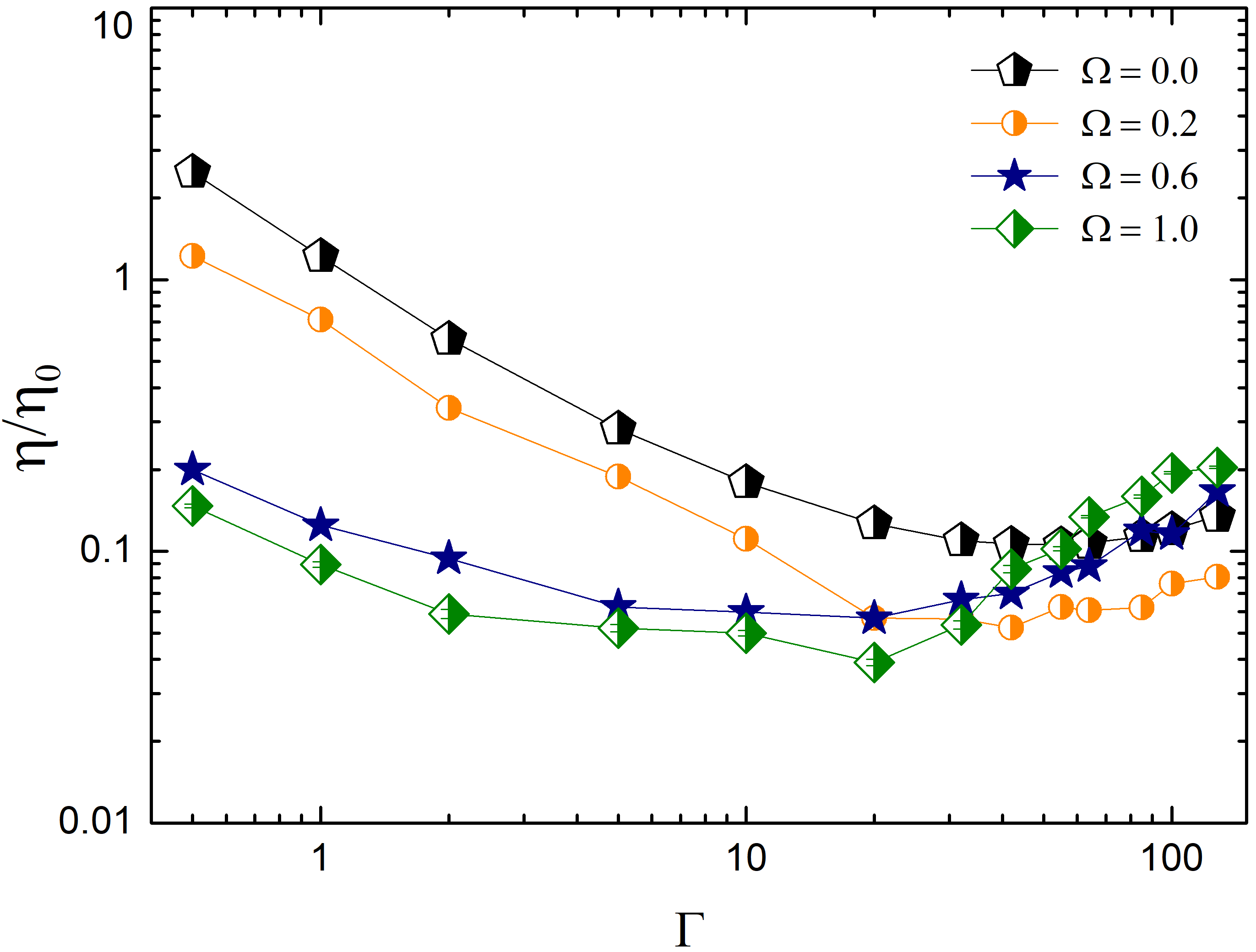}
        \centering
        \end{minipage}%
    \hspace{5mm} 
    \caption{Dependence of the shear viscosity on the coupling parameter $\Gamma$ for different values of the magnetic field strength $\Omega$.}
    \label{fig:3}
\end{figure*}

The figure \ref{fig:3} shows the dependence of the computed reduced shear viscosity $\eta/\eta_0$ 
on the coupling parameter $\Gamma$ for the magnetization parameters $\Omega = 0.0$, $\Omega =0.2 $, $\Omega = 0.6$, and $\Omega = 1.0$. The shear viscosity in units of $\eta_0=mn\omega_pa^2$ is computed. In the unmagnetized case ($\Omega = 0$), the viscosity exhibits a well-known non-monotonic dependence on $\Gamma$: at first it decreases with the increase in the coupling parameter, reaches a minimum at intermediate coupling, and increases again in the strongly coupled regime due to enhanced correlations. The presence of a magnetic field reduces viscosity in the weak-coupling regime. In contrast, at larger $\Gamma$ values (closer to $\Gamma=100$), the magnetization yields higher viscosity. In general, a magnetic field results in the shift of the minimum of the shear viscosity dependence on $\Gamma$ toward smaller $\Gamma$ values. These effects become more pronounced with increasing $\Omega$. 

The reduced shear viscosity $\eta/\eta_0$ and the corresponding root-mean-square deviations $\langle \sigma \rangle$ for different values of the coupling parameter $\Gamma$ and the magnetization parameter $\Omega$, covering regimes from weak to strong coupling and from zero to moderate magnetization, are given in Table~1. The root-mean-square deviation $\langle \sigma \rangle$ remains small throughout the range of investigated parameters, indicating a good statistical convergence of the simulations and high reliability of the shear viscosity values obtained.

\subsection{Diffusion coefficient}

\begin{figure} 
\centering
\includegraphics[width=1.0\textwidth]{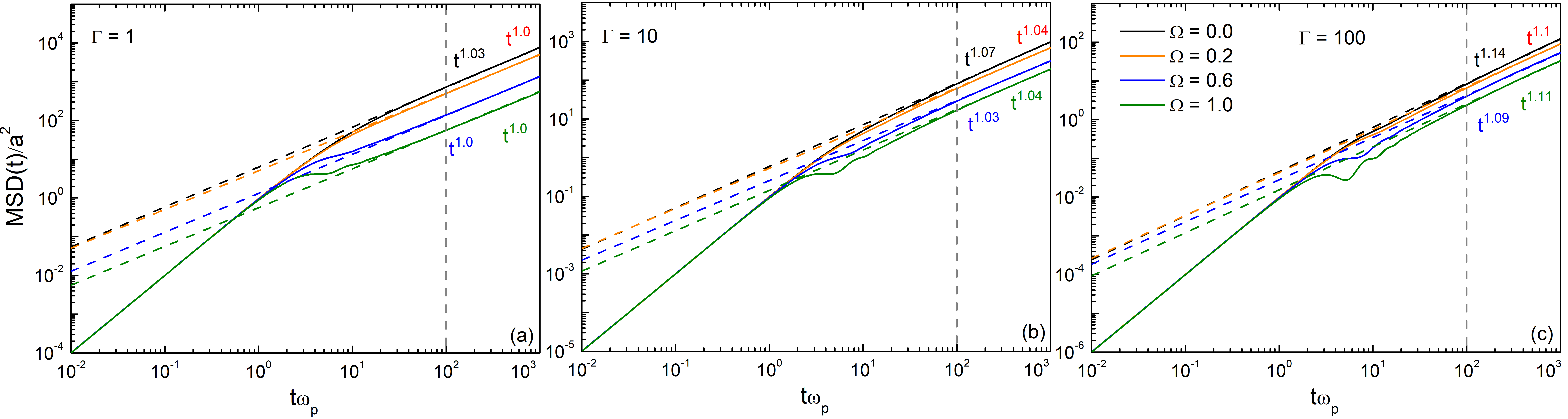}
\caption{  Mean-squared displacements  at different values of the magnetization parameter $\Omega$ for a) $\Gamma = 1$, b) $\Gamma = 10$, and c) $\Gamma = 100$.}
\label{fig:4}
\end{figure}

In figure \ref{fig:4}, the MSD as a function of normalized time $t \omega_p$ at different values of  $\Omega $ for (a) $\Gamma = 1$, (b) $\Gamma = 10$, and (c) $\Gamma = 100$ are presented using solid lines. The dashed lines show power-law fits, $\text{MSD}(t) = b\, t^\alpha$, in the long-time regime, obtained from data for $100\leq t\omega_p \leq 1000$. Increasing $\Omega$ or $\Gamma$  reduces MSD values. The considered long-time MSD values provide the diffusion parameter $\alpha$ close to normal diffusion ($\alpha \approx 1$) with weak superdiffusive behavior at higher $\Gamma$. Using the obtained MSD data, the coefficient $D_\alpha$ is estimated using  Eq.~\ref{eq:Diff} over the range $100 \le t\omega_p \le 1000$, corresponding to experimentally accessible time scales~\cite{Hartmann2}. 

\begin{figure*}[ht!]
    \centering
    \begin{minipage}{.5\textwidth}
        \includegraphics[width=\linewidth]{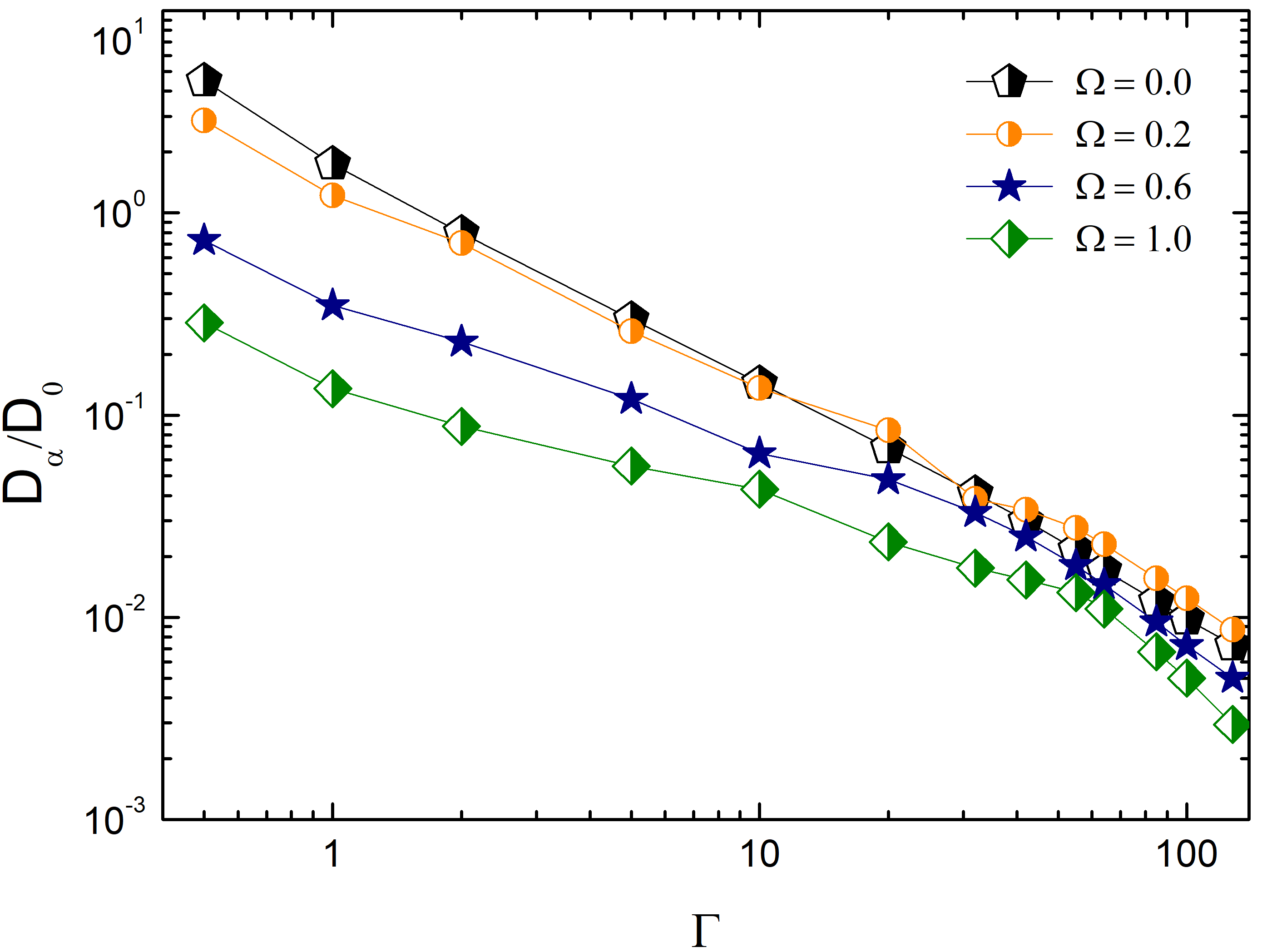}
        \centering
    \end{minipage}
    \caption{Dependence of the diffusion coefficient on the coupling parameter $\Gamma$ for different values of the magnetization parameter $\Omega$.}
    \label{fig:5}
\end{figure*}
\begin{figure} 
\centering
\includegraphics[width=0.9\textwidth]{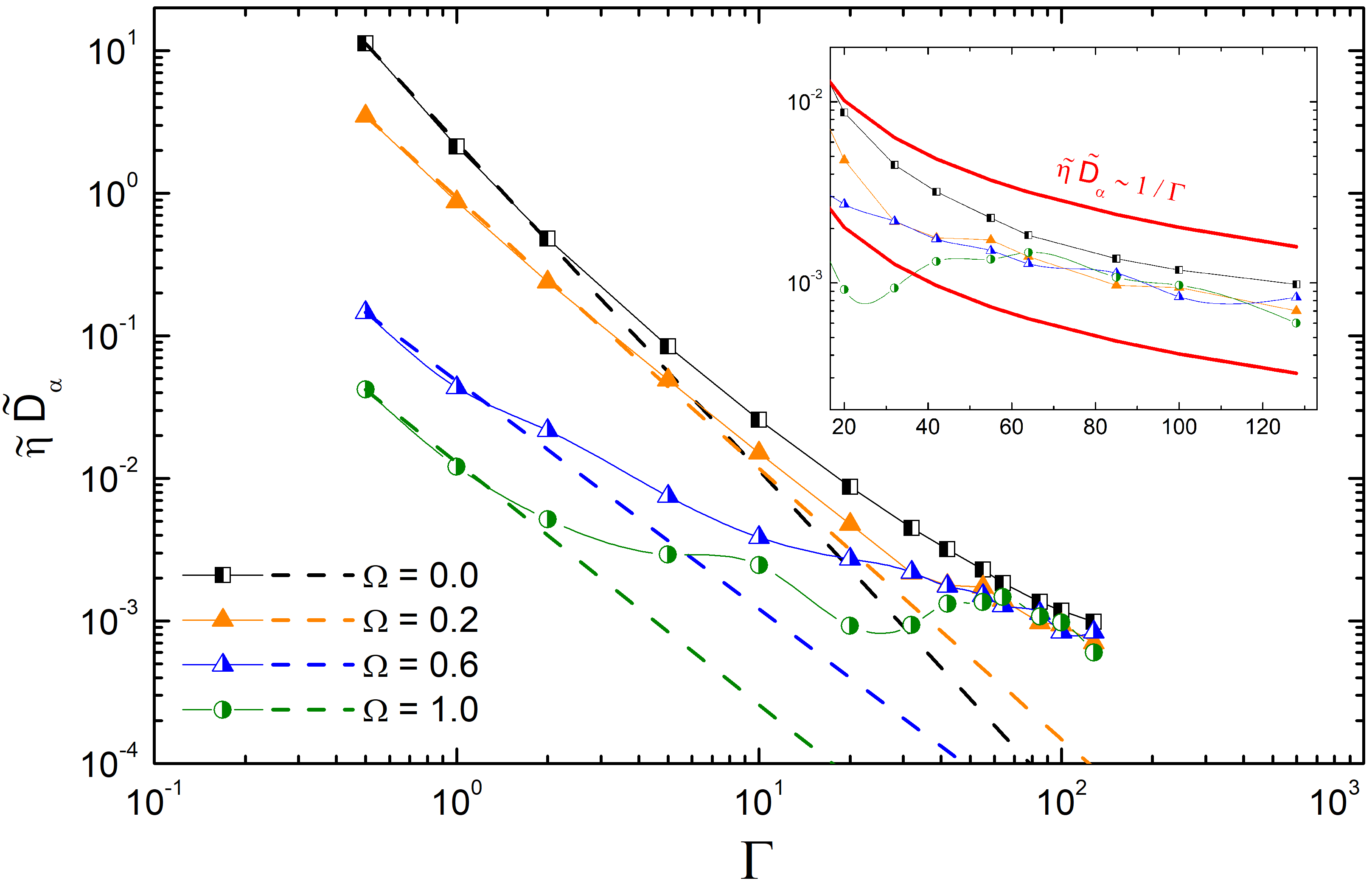}
\caption{Dependence of the product $\tilde{\eta}\tilde{D}_{\alpha}$ of the reduced shear viscosity $\tilde{\eta}=\eta/\eta_0$ and the reduced diffusion coefficient $\tilde{D}_{\alpha}=D_{\alpha}/D_{0}$  on the coupling parameter $\Gamma$ for different values of the magnetic field strength $\Omega$.}
\label{fig:6}
\end{figure}

In Figure \ref{fig:5}, we depict the dependence of the diffusion coefficient on the coupling parameter at different considered levels of magnetization. We compute the diffusion coefficient in units of $D_0=a^2\omega_p$.  
As the coupling parameter increases, the diffusion coefficient decreases monotonically for all considered magnetic-field strengths, indicating a gradual suppression of diffusive transport with increasing interparticle correlations. For  $\Gamma\lesssim40$, the magnetic field results in lower diffusion coefficients, demonstrating the additional constraint imposed by the magnetic field on particle motion.
In the strong-coupling regime, the separation between the curves corresponding to different values of $\Omega$ becomes smaller, indicating that the influence of the magnetic field on diffusion is reduced at high coupling strengths. For $\Gamma\gtrsim 40$ and $\Omega=0.2$, we observe an increase in the diffusion coefficient value compared to the case without a magnetic field, indicating competing effects of particle correlations and cyclotron motion at $\Omega\sim 0.1$. 
Indeed, one might expect the correlations to tend to locate particles in a local potential minima created by other charged particles, while cyclotron motion pushes particles outside of the local potential minima if the cyclotron radius is large enough. This effect disappears at large considered  $\Omega$ values as the cyclotron radius reduces with the increase in the magnetization degree. 

The computed values of the diffusion coefficient $D_{\alpha}$ and the corresponding estimated standard errors (as calculated over the range $100 \le t\omega_p \le 1000$) for different coupling parameters $\Gamma$ and magnetization parameters $\Omega$ are provided in Table~2.  Standard errors are defined as the square roots of the diagonal elements of the covariance matrix obtained from the used fit.

\subsection{Stokes--Einstein type relation under strong magnetic fields}

Figure \ref{fig:6} shows the dependence of the product $\tilde{\eta}\tilde{D}_{\alpha}$ of the reduced shear viscosity $\tilde{\eta}=\eta/\eta_0$ and the diffusion coefficient $\tilde{D}_{\alpha}=D_{\alpha}/D_{0}$  on the coupling parameter $\Gamma$ for magnetization parameters $\Omega = 0.0$, $\Omega =0.2 $, $\Omega = 0.6$, and $\Omega = 1.0$. For all considered magnetic field strengths, $\tilde{\eta}\tilde{D}_{\alpha}$ decreases monotonically with increasing $\Gamma$. Increasing magnetization systematically shifts the curves downward and alters their slopes, demonstrating a strong magnetic-field influence on the relationship between viscous and diffusive transport. 

From Figure \ref{fig:6}, we see that in the weak and intermediate coupling regime ($\Gamma \lesssim 30$), the influence of the magnetic field on the product $\tilde{\eta}\tilde{D}_{\alpha}$, where curves for different values of $\Omega$ are clearly separated, and one can identify the depends of the scaling behavior on the magnetization. 
At small coupling values $\Gamma\lesssim 10$, one can observe a nearly linear dependence of  $\log_{10}\left(\tilde{\eta}\tilde{D}_{\alpha}\right)$ on $\log_{10}\left(\Gamma\right)$ (as depicted using dashed lines).
We parametrize the corresponding fits (dashed lines) by the following equations:
\begin{equation}\label{eq:etaD_Omega0}
\tilde{\eta} \tilde{D}_{\alpha} 
= \frac{\Delta }{\Gamma^{c}}\,
\end{equation}
where $\Delta = 2.28 , c=2.3$ for $\Omega = 0$; $\Delta = 0.94, c=1.9$ for $\Omega = 0.2$; $ \Delta = 0.05, c=1.6$ for $\Omega = 0.6$; $\Delta = 0.012 , c=1.7$ for $\Omega = 1.0$. Note that the standard Stokes--Einstein relation requires $c=1$. 

At larger values of the coupling parameter, the dependence of the product $\tilde{\eta}\tilde{D}_{\alpha}$ on the coupling parameter becomes more complex and cannot be described by a power low.  These observation indicate a pronounced violation of the Stokes--Einstein relation (with $c=1$) over a wide range of coupling parameters. This deviation can be attributed to the strong interparticle correlations, collective dynamic, and transient caging effects, which influence viscous and diffusive transport in different ways. When a magnetic field is applied, deviations from the Stokes--Einstein relation persist, with the increasing the magnetic-field strength leading to a reduction in the power-law exponent $c$ compared to the case without magnetic field. 

An interesting observation is that for $\Gamma \gtrsim 50$, the product $\tilde{\eta}\tilde{D}{\alpha}$ exhibits a nearly identical dependence on the coupling parameter within the range $60 \lesssim \Gamma \lesssim 120$. This behavior can be approximately described by the standard Stokes--Einstein relation, $\tilde{\eta}\tilde{D}{\alpha} \sim 1/\Gamma$ (i.e., with $c = 1$). This trend is illustrated in the enlarged view (inset) of Figure \ref{fig:6}, where $\tilde{\eta}\tilde{D}_{\alpha}$ is seen to approximately follow the Stokes--Einstein scaling with $\Gamma$. The solid red lines, which bound the data points from above and below in the relevant range, are included as a guide to the eye. Notably, this behavior is observed for all considered values of the magnetization parameter $\Omega$, suggesting an approximate restoration of the Stokes--Einstein relation in the strong-coupling regime, largely independent of the external magnetic field. Without a magnetic field, a similar observation was previously reported for a two-dimensional Yukawa system with $88<\Gamma<124$\cite{PhysRevLett.96.015005}, which independently confirms our finding that the Stokes--Einstein relation is approximately valid at  $60 \lesssim \Gamma \lesssim 120$.

\section{Conclusions}\label{sec1}
We examined the relationship between shear viscosity and the diffusion coefficient, analogous to the Stokes-Einstein relation, for a two-dimensional Yukawa liquid under an external magnetic field over a wide range of coupling and magnetization parameters.  Our analysis reveals that the product \(\tilde{\eta}\tilde{D}_{\alpha}\) demonstrates a nonlinear dependence on the coupling parameter, diverging from the Stokes-Einstein relation under all explored coupling and magnetization conditions. At low coupling, specifically for \(\Gamma \lesssim 10\), this relationship can be described by a power law: \(\tilde{\eta}\tilde{D}_{\alpha} \sim 1/\Gamma^c\) with \(c > 1\). Conversely, within the range of \(60 \lesssim \Gamma \lesssim 120\), the standard Stokes-Einstein scaling, \(\tilde{\eta}\tilde{D}_{\alpha} \sim 1/\Gamma\), is roughly restored for all considered values of the magnetization parameter (\(0 \leq \Omega \leq 1\)), aligning with prior results obtained without a magnetic field \cite{PhysRevLett.96.015005}.

These results can be tested through experiments on two-dimensional quasi-magnetized rotating dusty plasmas \cite{Hartmann2}. Additionally, the calculated shear viscosity and diffusion coefficients serve as a valuable benchmark for evaluating fluid theories of charged systems in the presence of a magnetic field.

\begin{table}[h]
\centering
\caption{Dependence of $\eta/\eta_0$ and the root-mean-square deviation $\langle \sigma \rangle$ on $\Gamma$ for different values of $\Omega$.}
\label{tab:viscosity}
\renewcommand{\arraystretch}{1.2}
\begin{tabular}{c|cc|cc|cc|cc}
\hline
\multirow{2}{*}{$\Gamma$} 
& \multicolumn{8}{c}{$\Omega$} \\
\cline{2-9}
& \multicolumn{2}{c}{$\Omega = 0.0$}
& \multicolumn{2}{c}{$\Omega = 0.2$}
& \multicolumn{2}{c}{$\Omega = 0.6$}
& \multicolumn{2}{c}{$\Omega = 1.0$} \\
\cline{2-9}
& $\eta/\eta_0$ & $\langle \sigma \rangle$
& $\eta/\eta_0$ & $\langle \sigma \rangle$
& $\eta/\eta_0$ & $\langle \sigma \rangle$
& $\eta/\eta_0$ & $\langle \sigma \rangle$ \\
\hline
0.5  & 2.489   & 0.01     & 1.222   & 0.031      & 0.200   & 0.008      & 0.147   & 0.0022 \\
1    & 1.21834 & 0.0165      & 0.7141  & 0.0347      & 0.1248  & 0.0047      & 0.0895  & 0.0021 \\
2    & 0.5868  & 0.0089  & 0.3372  & 0.0111  & 0.0911  & 0.0035  & 0.0588  & 0.0023 \\
5    & 0.3047  & 0.0136  & 0.1883  & 0.0063  & 0.0624  & 0.0009  & 0.0551 & 0.0017 \\
10   & 0.17919      & 0.0043      & 0.111      & 0.0028     & 0.06     & 0.0009      & 0.056      & 0.0023 \\
20   & 0.1186  & 0.0019  & 0.0755  & 0.0025  & 0.0569  & 0.0008  & 0.0447 & 0.0008 \\
32   & 0.0981  & 0.0048  & 0.0725  & 0.0009 & 0.0664 & 0.001 & 0.0539 & 0.002 \\
42   & 0.1026  & 0.0051  & 0.0747  & 0.001 & 0.0704 & 0.0011 & 0.0861 & 0.0027 \\
55   & 0.104  & 0.0049  & 0.0845  & 0.0021 & 0.0838 & 0.0013 & 0.1021 & 0.0014 \\
64   & 0.0912  & 0.0022  & 0.0829  & 0.0012 & 0.0882 & 0.0018 & 0.1131 & 0.0016 \\
85   & 0.0963  & 0.0019  & 0.0944  & 0.0035 & 0.1191 & 0.0018 & 0.1561 & 0.0031 \\
100  & 0.1202      & 0.0057    & 0.076      & 0.0006    & 0.1159      & 0.0061      & 0.1931    & 0.0039 \\
128  & 0.1224  & 0.0013  & 0.1120  & 0.0041 & 0.1668 & 0.0024 & 0.2039 & 0.0019 \\
\hline
\end{tabular}
\end{table}

\begin{table}[h]
\centering
\caption{Dependence of the diffusion coefficient $D_{\alpha}/D_{0}$ and the corresponding estimated fitting errors (as calculated in the range $100 \le t\omega_p \le 1000$)  on the coupling parameter $\Gamma$ for different values of the magnetic field strength $\Omega$. Standard errors (SE) of the parameters are defined as the square roots of the diagonal elements of the covariance matrix obtained from the nonlinear fit.}
\label{tab:diffusion}
\renewcommand{\arraystretch}{1.2}
\begin{tabular}{c|cc|cc|cc|cc}
\hline
\multirow{2}{*}{$\Gamma$}
& \multicolumn{8}{c}{$\Omega$} \\
\cline{2-9}
& \multicolumn{2}{c}{$\Omega = 0.0$}
& \multicolumn{2}{c}{$\Omega = 0.2$}
& \multicolumn{2}{c}{$\Omega = 0.6$}
& \multicolumn{2}{c}{$\Omega = 1.0$} \\
\cline{2-9}
& $D_{\alpha}/D_{0}$ & SE
& $D_{\alpha}/D_{0}$ & SE
& $D_{\alpha}/D_{0}$ & SE
& $D_{\alpha}/D_{0}$ & SE \\
\hline
0.5  & 4.51   & 0.0420  & 2.862  & 0.01311  & 0.731  & 0.0071  & 0.286   & 0.0027 \\
1    & 1.745  & 0.0172  & 1.222  & 0.011  & 0.348  & 0.0018  & 0.135   & 0.0002 \\
2    & 0.794  & 0.0019 & 0.708  & 0.00215 & 0.23   & 0.001  & 0.088   & 0.00019 \\
5    & 0.298  & 0.0013 & 0.26   & 0.00054 & 0.12   & 0.0009  & 0.056   & 0.00028 \\
10   & 0.144  & 0.00097 & 0.136  & 0.00062 & 0.0645 & 0.00042  & 0.043   & 0.00019 \\
20   & 0.0689 & 0.00066 & 0.084  & 0.00019 & 0.048  & 0.00012  & 0.0236  & 0.00012 \\
32   & 0.041  & 0.00039 & 0.0386 & 0.00018 & 0.033  & 0.00026  & 0.0175  & 0.00014 \\
42   & 0.03   & 0.00024 & 0.034  & 0.00015 & 0.025  & 0.00019  & 0.01531 & 0.00011 \\
55   & 0.0216 & 0.000051 & 0.0277 & 0.00008 & 0.018  & 0.00016  & 0.01325 & 0.00009 \\
64   & 0.0171 & 0.00015 & 0.023  & 0.000098 & 0.0145 & 0.00011  & 0.011   & 0.00007 \\
85   & 0.012  & 0.00009 & 0.0156 & 0.000042 & 0.0095 & 0.000084 & 0.00675 & 0.000045 \\
100  & 0.0098 & 0.000103 & 0.0124 & 0.000036 & 0.00725 & 0.000038 & 0.005   & 0.000027 \\
128  & 0.0073 & 0.000068 & 0.0087 & 0.00001 & 0.005  & 0.000019 & 0.00295 & 0.000012 \\
\hline
\end{tabular}
\end{table}

\section*{Acknowledgments}
This research is funded by the Science Committee of the Ministry of Education and Science of the Republic of Kazakhstan (Grant AP23488907).

\section*{Data availability}
The datasets used and/or analyzed during the current study are available from the corresponding author on reasonable request.

 \bibliographystyle{apsrev4-1}
 \bibliography{ref}
\end{document}